\documentstyle[floats,aps,prl,epsf,twocolumn]{revtex}
\begin{document}
\draft
\wideabs{
\title{Luttinger-liquid-like transport in long InSb nanowires\\}

\author{S.V.~Zaitsev-Zotov$^1$, Yu.A.~Kumzerov$^2$, 
Yu.A.~Firsov$^2$, P.~Monceau$^3$\\}

\address{$^1$Institute of Radioengineering and 
Electronics of Russian Academy of Sciences, Mokhovaya 11, 
103907 Moscow, Russia\\}

\address{$^2$A.F.~Ioffe Physical-Technical Institute of 
Russian Academy of Sciences, Polytekhnicheskaya  26, 
194021 Sankt-Petersburg, Russia\\}

\address{$^3$Centre de Recherches sur Les Tr\`{e}s Basses 
Temp\'{e}ratures, C.N.R.S., B.P. 166, 38042\\
Grenoble C\'{e}dex 9, France\\}

\maketitle

\begin{abstract}
Long nanowires of degenerate semiconductor InSb in asbestos matrix 
(wire diameter is around 50 \AA , length 0.1 - 1~mm) were prepared. 
Electrical conduction of these nanowires is studied over a temperature 
range 1.5 - 350~K. It is found that a zero-field electrical conduction is 
a power function of the temperature $G\propto T^\alpha$ with the typical 
exponent $\alpha \approx 4$. Current-voltage characteristics of such 
nanowires are found to be nonlinear and at sufficiently low temperatures 
follows the power law $I\propto V^\beta$. 
It is shown that the electrical conduction of these nanowires cannot be 
accounted for in terms of ordinary single-electron theories and exhibits 
features expected for impure Luttinger liquid. For a simple approximation
of impure LL as a pure one broken into drops by weak 
links, the estimated weak-link density is around $10^3-10^4$ per cm.
\end{abstract}

\pacs{PACs numbers: 73.20.Dx, 73.50.Fq,73.61.Ey}
}


Electron-electron correlation effects being negligible in three-dimensional 
case play the dominant role in one dimension. As a result, the physical 
properties of a one-dimensional (1D) metal are expected to be dramatically 
different from properties of usual metals with a Fermi liquid of electrons. 
One of the main 
consequences of the electron-electron Coulomb repulsive interaction is 
the decrease of the energy density of states around the Fermi energy. 
The resulting electron state depends on details of electron-electron 
interaction. In the absence of the long-range interaction an 1D electron 
liquid (so-called ``Luttinger liquid''~\cite{Haldane} (LL)) is formed,  
whereas long-range Coulomb interaction leads to 1D
Wigner crystal~\cite{Schultz}. Transport properties of 1D electron systems 
are the subject of very high interest. 
It was shown that in Coulomb blockade systems the  
tunnelling transparency of a barrier vanishes due
to electron-electron interaction resulting to the power-law 
zero-bias anomaly for the conduction~\cite{GGJPS} 
(see also~\cite{MYG}). For the particular case of 
LL similar result was also obtained 
by more strong methods~\cite{KF1,KF2,FN1,Wan,MG1}. 
A role of impurity scattering in the presence of 
short-range~\cite{FN1,GS} and long-range~\cite{MG1,MG2} 
interactions was also intensively studied.

Luttinger-liquid-like behaviour for tunnelling 
into fractional Hall edge 
states predicted in~\cite{Wan} was recently observed in a 
GaAs-Al$_{0.1}$Ga$_{0.9}$As 
heterostructure in quantum Hall regime with the filling 
factor 1/3~\cite{CPW}. However each of these edge states have 
properties of so-called ``chiral Luttinger liquid'' but not those of
LL. They are similar but not identical. 
Luttinger-liquid behaviour was predicted also for
carbon nanotubes~\cite{EGcarbon,KBF}. Small 
(below one order of magnitude) voltage and 
temperature variations of the conductance of 
single-walled nanotubes consistent with 
the expectations for LL were observed experimentally~\cite{carbon}.

In the present paper 
we report an electric conduction of InSb nanowires in an asbestos 
matrix. It has been found that it 
follows the power law over 5 orders of conduction variation and can
be described as conduction of a pure LL broken into drops by weak 
links, the estimated weak-link density being around $10^3-10^4$ per cm.

Natural asbestos is a mineral, which is very convenient for nanowire 
preparation. It represents itself a crystal-like package of long fiber-like 
tubes of roughly 300~\AA \ in diameter with a central hole of approximately 
50~\AA \ in diameter \cite{Asbestos1,Asbestos2}. Almost any material at sufficiently 
high pressure and temperature can fill the holes. As the distance between  
nanowires is large, nanowires are independent of each other, in 
contrast to numerous quasi-one dimensional conductors where  
overlapping of wave functions of electrons on metal chains leads to 
3D-ordered charge-density waves. 
Superconducting properties of In, Sn and Hg 
nanowires~\cite{AsbestosSC1,AsbestosSC2,AsbestosSC3}, 
weak localisation phenomena \cite{AsbestosWL} and the first order 
melting transition in Hg nanowires in asbestos~\cite{AsbestosMelting} 
were studied earlier. 

In the present case of InSb as a nanowire 
material the asbestos was filled under the pressure 15~kbar at 
temperature 550~$^\circ$C. Preliminary pumping was undertaken to
evacuate gases absorbes by asbestos. 
In contrast to many other filling materials 
\cite{AsbestosSC1,AsbestosSC2,AsbestosSC3,AsbestosWL,AsbestosMelting} 
InSb nanowires 
were found to be stable at room temperature even after removing 
the pressure.

Samples were cut from big crystals of asbestos filled by InSb. 
Typical sizes of studied samples are 0.01-0.1~mm for their width 
and thickness and 0.1-1~mm for the length along the fibers of asbestos. 
Thus each sample contains $10^5-10^7$ nanowires. We found that 
in most cases such samples have also longitudinal cracks filled 
by InSb. To eliminate the effect of such cracks the samples were 
etched in 1:1 mixture of HCl and HNO$_3$ during 10-40 minutes. 
As such a procedure removes also InSb from the ends of asbestos 
filaments, the samples were shortened after etching to remove 
empty ends. 

In most cases the current terminals were made by vacuum 
deposition of indium, in few cases In-Ga amalgam was used 
for contact preparation. 
We used two-terminal measurement 
technique for measurements of electrical conduction of both 
pure asbestos and InSb in asbestos.
No difference was found between conduction data of 
samples with both types of contacts, but the 
vacuum-deposited ones were found to be much more stable. 
Electrical conduction of 
bulk InSb samples was measured by both 2 and 4-terminal 
technique. No substantial contact effects was observed.
Measurements of temperature dependencies of 
conductance and current-voltage characteristics (I-V curves) 
were done in the voltage-controlled regime. Electrometric 
amplifier U5-11 and electrometer Keithley 617 were used for 
measurements of currents down to  $10^{-15}$~A.

We prepared and tested the room-temperature conductance of 
more that 80 samples of InSb in asbestos. Around 50 of them 
were chosen for detailed investigation of temperature and electric
field conduction variation. 
For comparison we also measured the electrical conduction of 
InSb in multyfilamentary glass (vycor) with typical pore diameter 
around 70 \AA , of bulk pieces of InSb extracted from cracks, as 
well as the conductance of empty asbestos.  

We found that all measured samples can be divided into three 
groups in accordance with temperature evolution of their conductance. 
The first group consists of samples with 
$G({\rm 300~K})/G({\rm 4.2~K})\lesssim 100$. Practically all non-etched 
samples with vacuum-deposited contacts belong to this group. 
For such samples a rapid initial decrease of $G$ upon cooling 
saturates at some temperature so that their low-temperature $G(T)$ is 
much similar with $G(T)$ of a bulk InSb. 
We concluded therefore that in such samples the dominant contribution 
into low-temperature conduction is provided by InSb-filled 
longitudinal cracks
of asbestos. The second group is formed by samples whose 
conductance has strong temperature dependence down to the 
lowest temperature of measurements (1.5~K). From our point of 
view the conductance of these samples is provided by InSb 
nanowires in an asbestos matrix. The physical properties of such 
samples are the main subject of the present paper. In addition, 
some samples have very low room-temperature conductance 
(around $10^{-12}$~Ohm$^{-1}$) and very strong temperature 
dependence of conductance which goes out of the range of 
the present measurements ($\sim 10^{-15}$~Ohm$^{-1}$) at temperatures 
below 100 - 200 K. We believe that InSb nanowires in such 
samples are not continuous. The properties of such samples 
are out of the scope of the present paper. In addition, the 
conductance of some samples was found to be not stable 
and switches between two or more different conducting 
states belonging to one of the conduction groups described above. 
Such samples 
do not provide any new information, and the temperature 
evolution of their conductance will not be described below. 

Current-voltage characteristics of samples 
of the second group are highly nonlinear 
in entire temperature range studied. We also found that application 
of a sufficiently large voltage may irreversibly modify the shape 
of I-V curve, a result of such modification being dependent 
on a particular sample. Conduction of some samples was stable 
at voltages up to 30 V, but only 3 V was enough to irreversibly 
modify the conduction of other samples. Both irreversible increase 
and decrease of conduction were observed. The physical reasons for such 
irreversible behaviour may be related with electromigration of 
impurities. The data reported below were obtained for the 
reversible part of I-V curves. 

\begin{figure}
\vskip -1cm
\epsfysize=15cm
\centerline{\epsffile{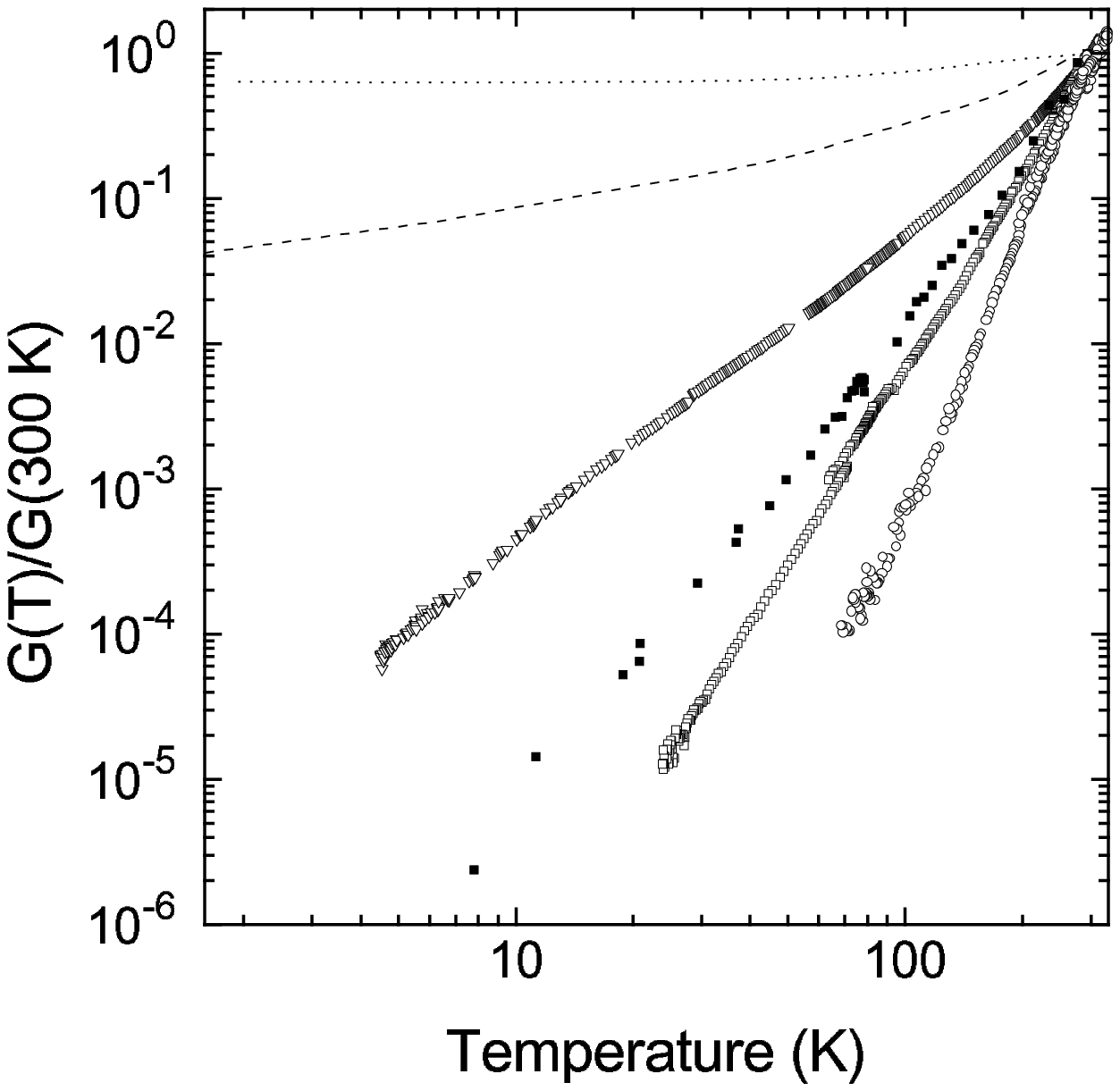}}
\vskip -6.5cm
\caption{Log-log plot of temperature variation of the linear 
conductance of InSb nanowires. Dashed line shows the 
conductance of a bulk InSb extracted from a crack in asbestos. 
Dotted line shows the conductance of InSb in vycor.}
\label{RT}
\end{figure}

Fig.~\ref{RT} shows typical dependencies of the linear conductance 
$G$ as a function of temperature for a representative set of 
samples of InSb-filled asbestos. The typical room-temperature resistance 
of such samples is 10 kOhm - 10 MOhm. $G(T)$ of InSb from extracted from
a crack in asbestos  (dashed line at Fig.~\ref{RT}) and InSb in vycor 
(dotted line at Fig.~\ref{RT}) are shown on the same plot for 
comparison. First of all, $G(T)$ of all samples was at least three orders of 
magnitude smaller than the conduction of a bulk InSb from a crack 
multiplied by the filling factor of asbestos $(50$~\AA$/300$~\AA$)^2$ and 
conduction of InSb in vycor. Secondly, the temperature dependence of 
their conductance was much stronger. The most remarkable 
feature  is approximately linear shape of $G(T)$ in log-log scale 
(see Fig.~\ref{RT}). It means that $G(T)\propto T^\alpha$. A large 
group of studied samples (six samples) has $\alpha = 4\pm 0.5$. 
Apart of this group one sample has $\alpha \approx 2$, and another 
one has $\alpha \approx 7$. 

\begin{figure}
\vskip -1cm
\epsfysize=11cm
\centerline{\epsffile{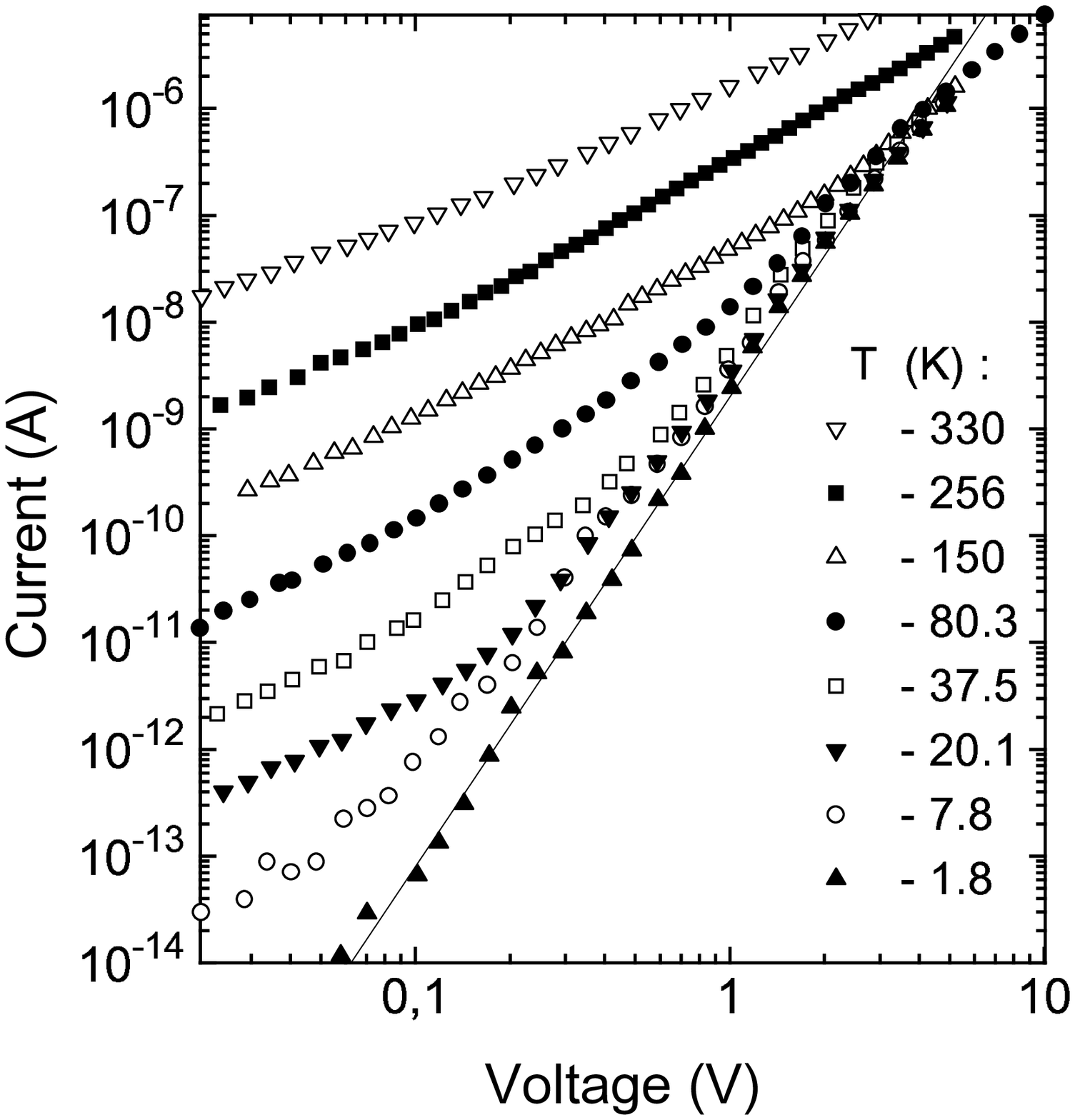}}
\vskip -3cm
\caption{Typical temperature set of I-V curves for a sample 
of  InSb in asbestos. Solid line shows $I\propto V^{4.4}$.
Sample length is 0.2~mm.}
\label{IV}
\end{figure}

Fig.~\ref{IV} gives a typical temperature set of I-V curves. The 
I-V curves are nonlinear at any temperature. The nonlinearity is 
getting more pronounced with lowering temperature. 
The power law, $I\propto V^{\beta+1}$,  with $\beta \approx 4.4$ 
fits the low-temperature data over 7 orders of 
the current variation. Other samples reveal  similar behavior 
with $2\lesssim\beta\lesssim 6$. Only a small nonlinearity (10\% )
was observed for InSb in vycor at $T = 4.2$~K  for electric fields 
$E<10$~V/cm .

The doping level of InSb from cracks in asbestos is expected to 
be close to one of InSb in nanowires. The weak temperature 
dependence of conduction of a bulk InSb indicates that InSb 
nanowires are  made of a heavy-doped degenerate  InSb with 
the carrier concentration $n$ above $10^{17}$~cm$^{-3}$. Qualitative 
difference between the temperature dependencies of the 
conductance of InSb in vycor glass and InSb nanowires indicates 
dramatic changes in the physical properties of InSb nanowires 
due 1D character of electron states, rather than the finite-size effect only. 
For an InSb wire with a diameter 
$d=50$~\AA\ and the current carrier's effective mass 
$m^*=0.014-0.2m_e$ the energy separation between the first 
and the second quantum levels  
$\Delta E=(\pi \hbar )^2 /2m^*d^2= 800 - 20000{\rm\ K}\gg T$ for 
the examined temperature range. So the electron band 
structure of 50~\AA\ InSb nanowires is essentially one-dimensional. 
Other details of the band structure are not very certain. 
In particular, the Fermi energy $E_F=(\pi \hbar nd^2)^2/8m^*$ 
depends on uncertain parameters $m^*$ and $n$. 
If $n\leq 2/d^3\sim 2\times 10^{19}$~cm$^{-3}$, 
then $E_F<\Delta E$, i.e. there is a single quantum conduction channel.  

Thus the studied samples consist of long quantum wires 
with one or few quantum conduction channels. 
The wires are made of degenerate semiconductor and 
contain large amount of impurities and defects.
Single-electron 1D variable-range hopping conduction
$G\propto\exp(-[T_0/T]^{1/2})$  apparently  does not  
fit the data. In principle, $I\propto V^\beta$ with 
$\beta = 1.5-3 $ are known for charge-limited injection currents 
in semiconductors 
and dielectrics~\cite{Injection}. Similar dependencies  were really 
observed by us for empty asbestos, as well as for asbestos with 
broken nanowires, with $\beta = 1.5-2.5$. As for InSb nanowires, 
$\beta >3$ observed for some samples is 
out of the range for the current-injection case~\cite{Injection}. 

$G\propto T^\alpha$ is predicted for tunnelling between two 
drops of pure LL~\cite{KF1,KF2,FN1}, for LL 
in a long wire with impurities~\cite{RA}, and
for 1D Wigner crystal~\cite{MG1}. 
Similarly, $I\propto V^\beta$ is predicted for 
tunnelling between two drops of pure LL~\cite{KF1,KF2,FN1}, for 
LL in a long wire with impurities~\cite{RA}, and 
for Coulomb blocade effect~\cite{GGJPS,MYG,MG2}.
However the dependence $G\propto \exp[-\nu \ln(T_0/T)^{3/2}]$ is predicted
for tunnelling of 1D Coulomb gas~\cite{NF}, Wigner crystal~\cite{GRS,FGS},
and for a pure LL in the presence of long-range Coulomb 
interaction~\cite{NF}. We suppose that in our multiwire samples
long-range interactions may be screened and the system behaves
itself more like LL. Tendency to Anderson localisation caused by
backward scattering on impurities is essentially reduced due to
repulsion of electrons~\cite{GS}. Hence in our opinion all properties
of the system are stipulated by two basic factors: i) by the power-like 
behaviour for ``pseudo-gap'' for density of states near the Fermi level,
ii) by the presence of weak links in each wire.

So the power laws for both $G(T)$ and $I(V)$ are predicted 
for a weak link between two pure LL~\cite{KF1,KF2}, as well 
as for impure LL~\cite{RA} approximated here as drops of LL 
connected through weak links in serial. For a sample consisting 
of $N$ identical quantum wires with $M$ independent weak links in each 
$G(T)$ and $I(V)$~\cite{KF1,KF2} can be rewritten as
\begin{equation}
G(T)=C_TN\frac{e^2}{\hbar}\left[\sum_{j=1}^M{\left(\frac{\tilde\varepsilon}{t_j}\right)^{2}}\right]^{-1}
\left(\frac{kT}{\tilde\varepsilon}\right)^{\alpha}
\label{eq:GT}
\end{equation}
and
\begin{equation}
I(V)= C_I{N} \frac{e^2}{\hbar} \left(\frac{\tilde\varepsilon}{e}\right)
\left[\sum_{j=1}^M{\left(\frac{\tilde\varepsilon}{t_j}\right)^{\frac{2}{\beta}}}\right]^{-\beta }
\left(\frac{eV}{\tilde\varepsilon}\right)^{\beta},
\label{eq:IV}
\end{equation}
where $C_T,C_I\sim 1$, $t_j$ is a measure of tunnelling transparency of $j$-th
weak link, $\tilde\varepsilon\sim E_F$ is an energy scale for LL, 
and $\alpha=2/g-2$ and $\beta=2/g^*-1$, where
$g$ and $g^*$ are bare and renormalized~\cite{RA} dimensionless 
constants ($g=g^*$ for a pure LL~\cite{KF1,KF2}).

Observed values are 2.3 and 2.3,  3.4 and 4.4,  4.5 and 3.8, 
4.6 and 3.0 for $\alpha$ and 
$\beta$ respectively. $\alpha=4$ corresponds
to $g=1/3$ that is considered as a typical value for LL theory~\cite{KF1,KF2}.
If $\beta =\alpha +1$, and all weak links are identical, 
$M$ can be estimated from Eqs.~(\ref{eq:GT}),(\ref{eq:IV}) without 
fitting parameters from comparison of I-V curves and $G(T)$.
For the sample with $\alpha=3.4$ one gets
$M\approx 20$, that corresponds to $10^3$ weak links/cm per a wire. 
Taking for other samples $\tilde\varepsilon\sim 10^3$~K 
and experimental values 
$I(5{\rm\ K},1{\rm \ V})=10^{-12},\ 10^{-9},\ 2\times10^{-13}$~A,
$G(5\ {\rm K})=10^{-15},\ 2\times 10^{-9},\ 10^{-14}$ 1/Ohm,  
one gets $M=2\times 10^3, 5\times 10^3, 5\times 10^3$ respectively.
The respective weak link density per a wire is around $10^4$ per cm.

Two kinds of contact effects are expected to affect 
the transport properties of quantum wires. 
The first one caused by electron reflection from 
the contact can be described as an additional contact 
resistance of the order of $h/2e^2=12.3$~kOhms~\cite{MS} per
each quantum wire. This contribution is negligible in the 
low-temperature region. The second one associated with interaction
of electrons with image charges induced in the contacts leads to potential 
redistribution along a nanowire~\cite{SS,EG}. In accordance
with~\cite{SS,EG} at sufficiently low temperatures 
even in the case of a quantum wire with a single barrier 
the major contribution comes from the barrier rather from the contact effect.
This is even more true for our multibarrier samples.

So LL theory provides plausible explanation of our experimental data. 
We would like to note that LL is mostly a convenient theoretical
approximation: long-range Coulomb interaction, backscattering and
interference effects yield modification of the theory. 
There are reasons to expect that more realistic description of InSb nanowires 
may be possible without substantial modification in $G(T)$ and $I(V)$.


Three of us (S.V.Z.-Z., Yu.A.K., and Yu.A.F) are grateful to 
C.R.T.B.T.-C.N.R.S. for hospitality during the initial stage of the 
research. We would like to thank S.N.~Artemenko and V.Ya.~Pokrovskii  
for their interest to this work, and A.M.~Nikitina for assistance in 
sample etching. This work has been supported by C.N.R.S. through 
jumelage 19 between C.R.T.B.T. and IRE RAS, the Region Rhone-Alpes
through the programme Tempra, the Russian Foundation for 
Basic Research (grants 97-02-18267, 98-02-16667), and MNTP 
``Physics of Solid State Nanostructures'' 96-2008 and 97-1052.

\end{document}